\documentclass{iau} 
\usepackage{graphicx}
\usepackage{subcaption} 

\title[JD 11.] 
{Can Bars Erode Cuspy Halos ?}

\author[Kataria et al]   
{Sandeep Kumar Kataria$^1$
 Mousumi Das$^2$ \and Stacy Mcgaugh$^3$}

\affiliation{$^1$Indian Institute of Astrophysics, Bangalore, 560034 and \\ Indian Institute of Science, Bangalore, 560012 ,
 \\ email: {\tt skkataria.iit@gmail.com} \\[\affilskip]
$^2$Indian Institute of Astrophysics, Bangalore, 560034  \\ [\affilskip]
$^3$Department of Astronomy, Case Western Reserve University, \\
Cleveland, OH 44106, USA 
}
\pubyear{2019}
\volume{353}  
\jname{Galactic Dynamics in the Era of Large Surveys}
\editors{M. Valluri, \& J. A. Sellwood  }
\begin{document}

\maketitle

\begin{abstract}
One of the major and widely known small scale problem with the Lambda CDM model of cosmology is the ``core-cusp" problem. In this study we investigate whether this problem can be resolved using bar instabilities. We see that all the initial bars are thin ($b/a<0.3$) in our simulations and the bar becomes thick ($b/a>0.3$) faster in the high resolution simulations. By increasing the resolution, we mean a larger number of disk particles. The thicker bars in the high resolution simulations transfer less angular momentum to the halo. Hence, we find that in the high resolution simulations it takes around 7 Gyr for the bar to remove inner dark matter cusp which is too long to be meaningful in galaxy evolution timescales. Physically, the reason is that as the  resolution increases, the bar buckles faster and becomes thicker much earlier on.  
\end{abstract}

\firstsection 
\section{Introduction}

Cosmological simulations confirm  that  dark  matter  halos  have
Navarro-Frenk-White type of density profiles which vary as $r^{-1}$ in the central region while varying as
$r^{-3}$ in the outer region \cite[(Navarro et al 1997)]{1997ApJ...490..493N}. This inner density profile which rises steeply in the central  region  is  often  called  a  “cusp”.   The  modeling  of HI  observations  of  dwarf  galaxies show that there is a discrepancy between the observed rotation curves and those calculated
theoretically, especially in the inner parts of the galaxies.  Galaxy rotation curves studies show that the dark matter profile varies as $r^{-\alpha}$ in the inner region where $\alpha = 0.2-0.4$, which
is also called “halo core” while it varies as $r^{-2}$ in the outer parts.  This is typically called the pseudo isothermal profile.  This discrepancy between observations and theoretical models of the inner profiles of galaxy halos is widely known as “Core-Cusp” problem of LCDM cosmology \cite[(de Blok 2010)]{2010AdAst2010E...5D}.

\begin{figure}
  \begin{subfigure}[b]{0.4\textwidth}
    \includegraphics[width=\textwidth]{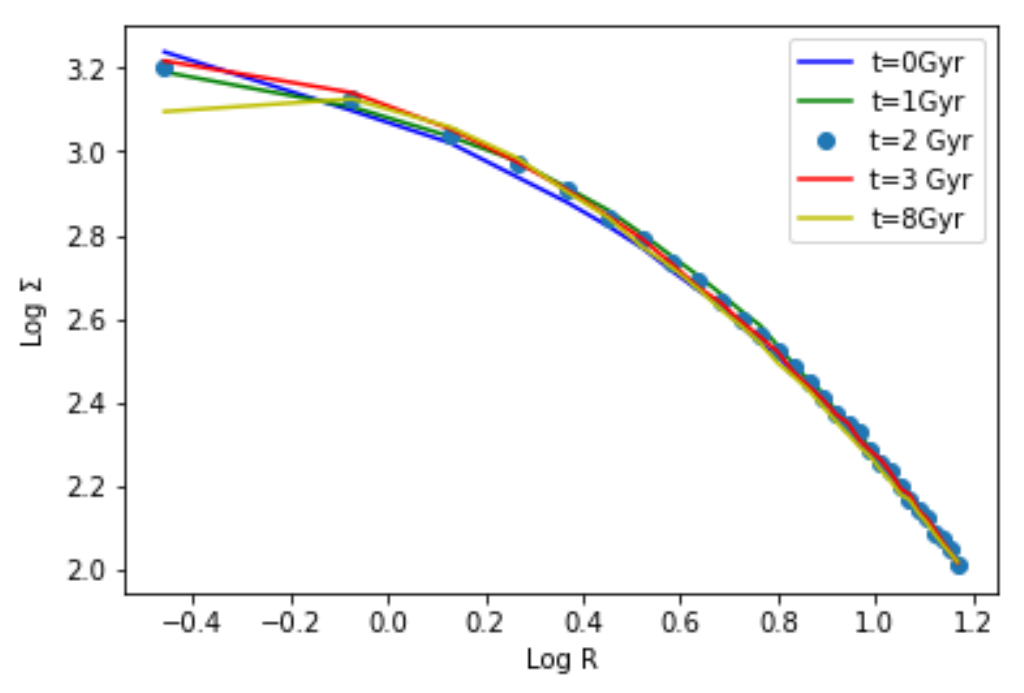}
    
  \end{subfigure}
  \begin{subfigure}[b]{0.4\textwidth}
    \includegraphics[width=\textwidth]{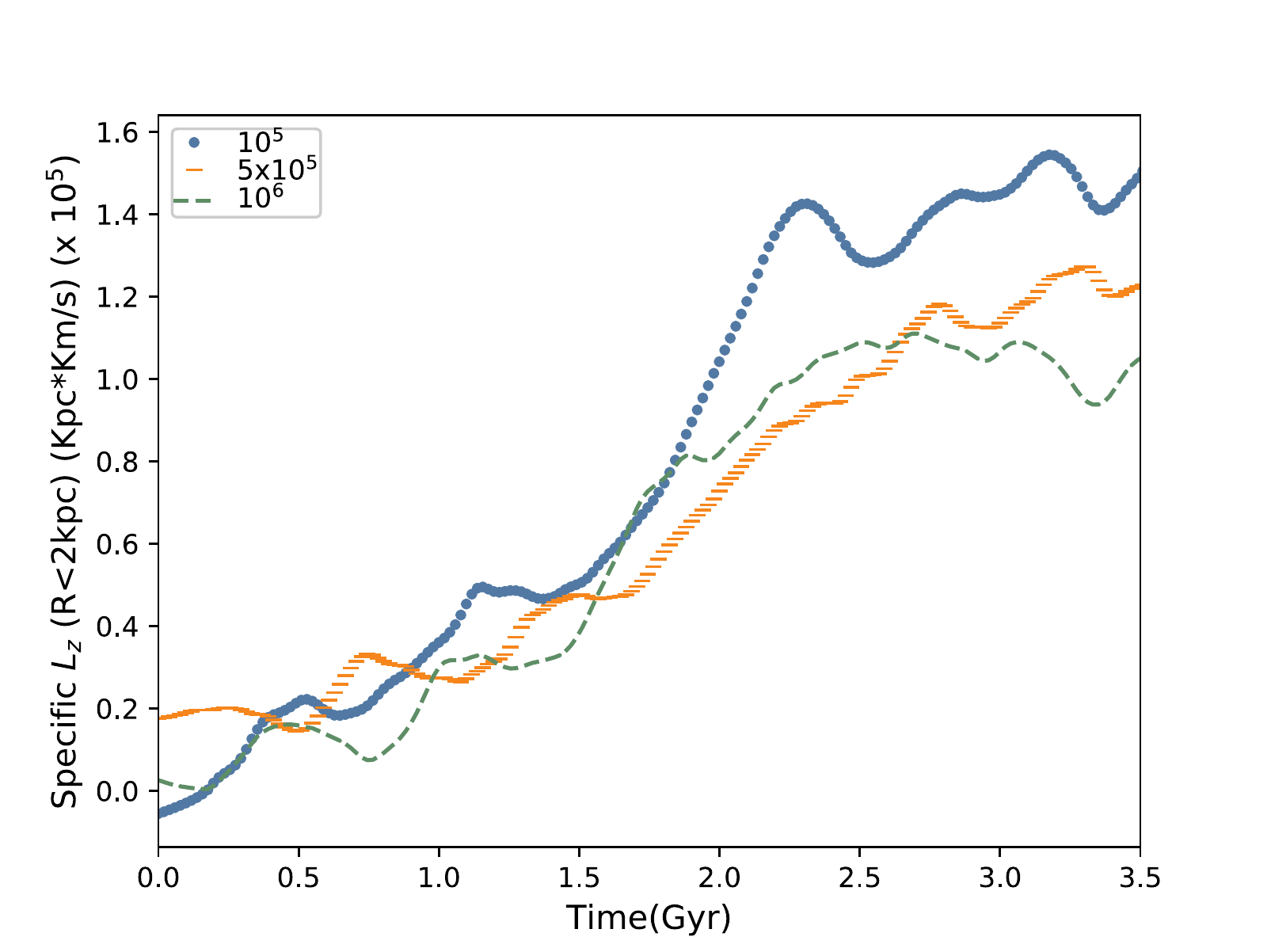}

  \end{subfigure}
  \caption{(a) Time evolution of halo profile of high resolution model, namely model H. (b) Angular momentum transfer from disk to inner dark matter within a sphere of 2 kpc.}
  \label{fig}
\end{figure}

In this study we use N-body simulation to examine whether bars can alter the initial cusps of dark matter halo profiles though angular momentum transport from disks to to halos. We have generated initial isolated disk models having exponential stellar disk and halo with Hernquist profiles using the GalIC code \cite[(Yurin \& Springel 2014)]{2014MNRAS.444...62Y}. We have generated three models namely L, M and H with increasing disk resolutions which means larger number of initial disk particles and not a shortening of the softening length. The number of disk particles are $10^5$, $5 \times 10^5$ and $10^6$ respectively. All the models are bar unstable within couple of Gyrs \cite[(Kataria \& Das 2018)]{2018MNRAS.475.1653K} after the isolated evolution using Gadget-2 \cite[(Springel 2005)]{2005MNRAS.364.1105S}. We find that as the resolution of simulation increases it leads to a different evolution of bar shapes. We find that all the initial bars are thin ($b/a<$0.3). As the bar evolves with time it transports angular momentum to inner halo through resonance interactions. We find that bar remains thin for low resolution simulations throughout it's evolution while it becomes thicker ($b/a>$0.3) quickly in high resolution simulations. The thickening is
probably due to the bars undergoing buckling instability and our simulations indicate that bars  buckle  faster  in  high  resolution  simulations  compared  to  low  resolution  ones. We have plotted the inner dark matter specific angular momentum variation in Fig. \ref{fig}b for all the simulations in order to check the rate of transfer of angular momentum to the inner dark matter halo component within  a 2 kpc radius. We see that as the resolution of the simulation increases (bar thickens faster), the total gain in angular momentum by the inner halo reduces.  As a result we see that bars are able to transform cuspy halo profile into cores within 4 Gyr of disk evolution in low resolution simulation while it doesn't in high resolution simulation. Instead we find that it takes around 7 Gyr of bar evolution in high resolution simulation to transform inner cusp into a core. We have plotted the evolution of dark matter halo profile in Fig. \ref{fig}a for high resolution simulation.

 \section{Summary} We test the role of bars in altering the inner dark matter halo profiles of disk galaxies. We start with cuspy halo models that have disks of increasing number of particles i.e. higher resolution. The disks in our models are bar unstable and form bars around 1.2 Gyr of isolated evolution. We find that the transfer of angular momentum from the disk to the halos by bars depends on the resolution of the simulations or timescale until the bar remains thin. For the high resolution simulation where initial thin bar buckles faster and becomes thick, it takes $\sim$7 Gyr of bar secular evolution(Fig. $\sim$ \ref{fig}b) to transform the halo cusp into a core. This suggests that timescales of cusp to core transition due to bars is too long to be meaningful for isolated galaxy evolution scenarios.

\textbf{Acknowledgement:} We gratefully acknowledge an IAU travel grant to attend the meeting
and IUSSTF grant JC-014/2017 for supporting the visit of S.K. to CWRU.

\end{document}